\documentclass[]{spie}  
\usepackage[]{graphicx}
\usepackage{amsmath,amssymb,mathtools,url}
\title{\Large \bf Scalar vortex coronagraph mask design and predicted performance } 
\usepackage{array}
\usepackage{hyperref}
\newcolumntype{P}[1]{>{\centering\arraybackslash}p{#1}}

\usepackage[table]{xcolor}

\author{
Garreth~Ruane\supit{a,b}, Dimitri~Mawet\supit{b,a}, A~J~Eldorado~Riggs\supit{a}, and Eugene~Serabyn\supit{a} \\
\supit{a}Jet Propulsion Laboratory, California Institute of Technology, 4800 Oak Grove Dr., \\Pasadena, CA 91109, USA\\
\supit{b}Department of Astronomy, California Institute of Technology, 1200 E. California Blvd., \\Pasadena, CA 91125, USA
}

\authorinfo{Send correspondence to gruane@jpl.nasa.gov}

\graphicspath{{Figures/}}
 
\begin{document} 
  \maketitle 

\begin{abstract}
Vortex coronagraphs are an attractive solution for imaging exoplanets with future space telescopes due to their relatively high throughput, large spectral bandwidth, and low sensitivity to low-order aberrations compared to other coronagraphs with similar inner working angles. Most of the vortex coronagraph mask development for space applications has focused on generating a polychromatic, vectorial, optical vortex using multiple layers of liquid crystal polymers. While this approach has been the most successful thus far, current fabrication processes achieve retardance errors of 0.1-1.0$^\circ$, which causes a nonnegligible fraction of the starlight to leak through the coronagraph. Circular polarizers are typically used to reject the stellar leakage reducing the throughput by a factor of two. Vector vortex masks also complicate wavefront control because they imprint conjugated phase ramps on the orthogonal circular polarization components, which may need to be split in order to properly sense and suppress the starlight. Scalar vortex masks can potentially circumvent these limitations by applying the same phase shift to all incident light regardless of the polarization state and thus have the potential to significantly improve the performance of vortex coronagraphs. We present scalar vortex coronagraph designs that make use of focal plane masks with multiple layers of dielectrics that (a) produce phase patterns that are relatively friendly to standard manufacturing processes and (b) achieve sufficient broadband starlight suppression, in theory, for imaging Earth-like planets with future space telescopes.
\end{abstract}



\section{INTRODUCTION}
\label{sec:intro} 

The direct imaging of exoplanets is a key science driver for next-generation space telescopes in the ultra-violet, optical, and near-infrared, such as the Habitable Exoplanet Explorer (HabEx)\cite{habexInterimReport} and the Large Ultra-violet, Optical, Infrared (LUVOIR) Surveyor\cite{LUVOIRinterimReport} mission concepts under consideration by the Astro2020 decadal survey. Coronagraph instruments for these missions\cite{Pueyo2017} have engineering requirements intended to enable the imaging of planets as small as Earth and as close to their stars as the inner habitable zone. Since the baseline coronagraph layouts are generally tailored to the telescope pupil, vortex coronagraphs\cite{Mawet2005,Foo2005,Ruane2018_JATIS} are a leading candidate for off-axis telescopes, where the secondary mirror doesn't obscure the telescope aperture\cite{Stark2019}. Vortex coronagraphs tend to give a desirable balance between several performance characteristics, including high throughput for planet light, a small inner working angle, relatively low sensitivity to tip/tilt errors and other low-order aberrations. Furthermore, depending on the design of the focal plane mask, vortex coronagraphs can potentially be used over a large spectral bandwidth. 

There are two main varieties of vortex coronagraphs: \textit{vectorial} and \textit{scalar}. The distinction is based on how the focal plane mask turns the starlight into an optical vortex. The former uses a half wave plate with spatially varying fast axis, which results in polarization dependent geometric phase shift\cite{Pancharatnam1956,Berry1984}, while the latter uses longitudinal phase delays. Most technical development thus far in this context has focused on manufacturing masks that generate a polychromatic, vectorial, optical vortex. In principle, vector phase masks can be manufactured using patterned liquid crystal\cite{Ganic2002,Marrucci2006,Mawet2009,Mawet2010a,Mawet2010b}, subwavelength gratings\cite{Biener2002,Mawet2005b,Niv2007}, photonic crystals\cite{Murakami2013}, metamaterials\cite{Genevet2012,Karimi2014}, or rotationally-symmetric prisms\cite{Bouchard2014}. However, the method that has provided the best coronagraph performance in the visible is using multiple layers of liquid crystal polymers to make up a spatially-variant diffractive waveplate\cite{Nersisyan2013,Serabyn2019}. 

Scalar vortex coronagraph approaches using spiral phase plates\cite{Swartzlander2008} or dispersion compensated holograms\cite{Errmann2013,Ruane2014} have not received as much attention in this context, despite the fact that the methods for producing a scalar optical vortex appear earlier in the literature\cite{Khonina1992,Heckenberg1992,Tidwell1993}. Here, we introduce scalar vortex concepts, investigate the coronagraph performance with a scalar vortex coronagraph based on the spiral phase plate and similar phase masks, and re-visit an early proposal by Swartzlander (2006)\cite{Swartzlander2006} to combine multiple materials to produce achromatic focal plane masks. 

\section{VORTEX CORONAGRAPH BASICS}

\begin{figure}
    \centering
    \includegraphics[width=0.8\linewidth]{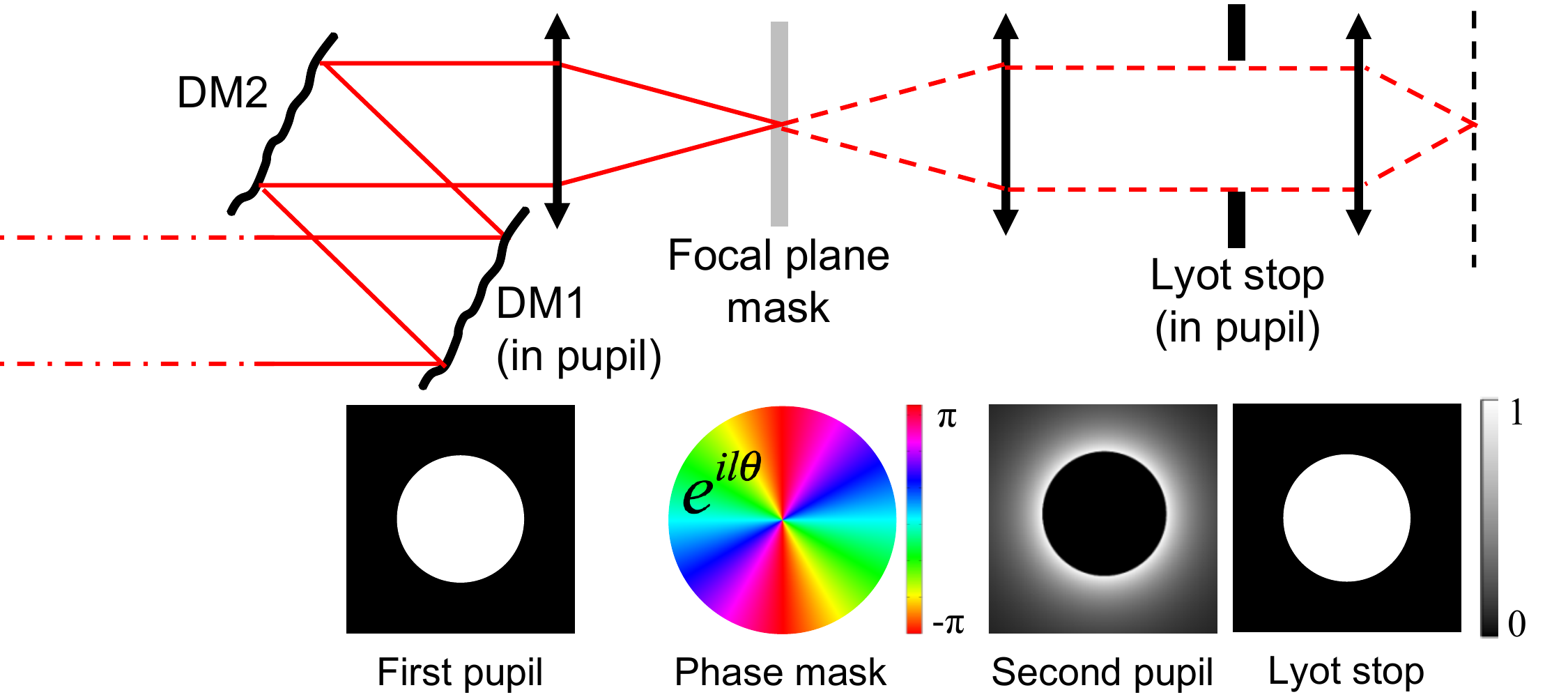}
    \caption{Schematic of a vortex coronagraph with two deformable mirrors (DM1 and DM2), a phase-only focal plane mask with transmission $t=\exp(il\theta)$, and circular Lyot stop. In the ideal case, the starlight is completely diffracted outside of the Lyot stop for all nonzero even values of $l$. Figure reproduced from Ruane et al. (2018)\cite{Ruane2018_JATIS}.}
    \label{fig:vc}
\end{figure}

A vortex coronagraph (see Fig.~\ref{fig:vc}) consists of a phase-only mask in the focal plane with complex transmission $t=\exp(il\theta)$, where $l$ is an even nonzero integer known as the charge. Starlight that is focused onto the center of the mask is diffracted outside of a circular aperture in the following pupil plane, known as the Lyot stop. The Lyot stop is typically undersized with respect to the full dark region in the second pupil. Two deformable mirrors upstream of the focal plane mask can be used to correct for wavefront errors and minimize the amount of starlight that appears within the desired field of view about the star where faint planets can be imaged. The value of $l$ trades throughput at small angular separations against the sensitivity of the coronagraph to tip/tilt errors and other low-order aberrations\cite{Mawet2010a}. For future space telescopes, $l\ge6$ will likely be desired to sufficiently relax low-order wavefront error requirements\cite{habexInterimReport,Ruane2018_JATIS}, especially in terms of differential polarization aberrations\cite{Breckinridge2018}. 

Although the underlying starlight suppression mechanism is identical, a number of practical details depend on whether the coronagraph uses a vector and scalar vortex mask; for instance, the mask manufacturing methods and the handling of polarization errors. The latter may have a significant impact on the overall design of the coronagraph instrument and performance. 

\section{FOCAL PLANE MASK DESIGNS AND PROPERTIES}

In this section, we briefly review the principle of a vector vortex mask and outline possible scalar alternatives.

\subsection{Vector vortex masks and their potential limitations}

A vector vortex mask is a half-wave plate with a spatially-variant fast-axis orientation $\chi(x,y)=l\theta/2$, which as derived in Appendix \ref{sec:appendixVectormasks}, may be described by the following Jones matrix in the circular polarization basis:
\begin{equation}
\mathbf{M}_\circlearrowright 
=c_V\left[\begin{matrix}
   0 & e^{il\theta}  \\
   e^{-il\theta} & 0  \\
\end{matrix} \right]
+ c_L\left[\begin{matrix}
   1 & 0  \\
   0 & 1  \\
\end{matrix} \right],
\label{eqn:vector}
\end{equation}
where $c_V$ and $c_L$ are constants\cite{Mawet2010a}. The first term indicates that a vortex phase ramp, $\exp(\pm il\theta)$, is applied to each polarization component, where the sign of the vortex ramp depends on the handedness of the incident circular polarization, and each beam transfers to the orthogonal polarization state. The second term is the stellar leakage component due to imperfect retardance; i.e. a part of the beam passes through the vector vortex without picking up a vortex phase whose fractional energy is given by $|c_L|^2\approx\epsilon^2/4$, where $\epsilon$ is the wavelength-dependent retardance error in radians, for $\epsilon\ll$1~rad. Equation~\ref{eqn:vector} reveals two potential limitations of vector vortex coronagraphs. 

The first potential limitation is the stellar leakage due to imperfect retardance, which resembles an additional Airy pattern in the final image plane. An important parameter for imaging exoplanet planets is the raw contrast near the star, or the ratio of the signal from a source at a given location to the signal from the star detected at that location. In order to achieve a raw contrast of $10^{-10}$, which is often used as a benchmark requirement for Earth-like exoplanets, $|c_L|^2$ should be $<10^{-8}$ such that diffraction rings from the leaked starlight are below $10^{-10}$. Therefore, the retardance requirement is $\epsilon<2\times10^{-4}$ radians, or 0.01$^\circ$, which is difficult to obtain for all wavelengths across typical astronomical passbands ($\Delta\lambda/\lambda\approx0.2$). Laboratory experiments aimed at achieving very high contrast with vector vortex masks filter the circular polarizations before and after the vortex mask such that only one of the two vortex beams is selected and the leakage term is blocked at the cost of 50\% in throughput\cite{Mawet2009,Serabyn2013}. To do this, a circular polarizer upstream of the vortex mask selects one circular polarization from the incident starlight and an orthogonal circular analyzer downstream blocks that polarization state. Since the beam with the desired phase switches to the orthogonal polarization at the vortex mask, it passes through the analyzer theoretically uninhibited. In this configuration, the raw contrast depends on the extinction ratio of the circular polarizer/analyzer pair (see Appendix \ref{sec:appendixVectormasks}). 

The second potential limitation is that, when using a vector vortex coronagraph in both polarizations simultaneously, each of the two circular polarization terms evolve differently through the optical system. That is, each pick up different aberrations before and after the vortex mask, and a different sign for the $e^{\pm il\theta}$ phase ramp at the vortex mask. When using wavefront control with deformable mirrors to optimize raw contrast, e.g. electric field conjugation (EFC)\cite{Giveon2007}, it may not be possible to find a single deformable mirror setting that makes the stellar intensity sufficiently small in the final image plane for both polarizations\cite{ACCESSreport}. Using the circular polarizer/analyzer pair also helps mitigate this by removing one of the two polarizations in addition to the stellar leakage term. 

While significant progress has been made in recent laboratory studies to achieve very high contrast with vector vortex coronagraphs\cite{Serabyn2013,Serabyn2019}, using scalar vortex masks is a viable alternative that would circumvent these two issues. However, scalar vortex masks have never been tested at high contrast (goal raw contrasts are typically between $10^{-10}$ and $10^{-8}$). In the following sections, we study the predicted performance of various scalar vortex designs that could be used instead of vector masks. 

\subsection{Scalar vortex optics}

The focal plane mask in a vortex coronagraph nominally has a complex transmission of the form $t=\exp(il\theta)$, where $l$ is ideally a nonzero, even integer. However, this is a subset of a larger family of azimuthal phase patterns\cite{Henault2018} that include four-quadrant\cite{Rouan2000}, multi-zone\cite{Murakami2008}, sinusoidal, staircase-like masks\cite{Lee2006,QiansenXie2008}, or any linear combination of even azimuthal modes $t=\sum_m C_{2m} e^{i2m\theta}$, where $t$ is phase-only function\cite{Jenkins2008,Ma2012,Hou2014}. In the following, we identify and compare practical designs for scalar vortex masks that emerge from this family.

\subsubsection{Reflective spiral phase mirror}

One way to create a scalar optical vortex is with directly machined spiral phase mirrors\cite{Campbell2012}. In reflection, the complex transmission is 
\begin{equation}
    t = \exp\left( i \frac{4\pi}{\lambda} h(\theta)\right),
\end{equation}
where $h(\theta)$ is the surface height of the mirror. In order to create an optical vortex of charge $l_0$ at the design wavelength, $\lambda_0$, the surface height is $h(\theta)=\lambda_0 l_0 \theta/(4\pi).$ Thus, the charge as a function of wavelength is then given by $l(\lambda) = l_0 \lambda_0/ \lambda$. By comparison, an ideal vortex coronagraph has a constant charge as a function of wavelength.

\begin{figure}[t]
    \centering
    \includegraphics{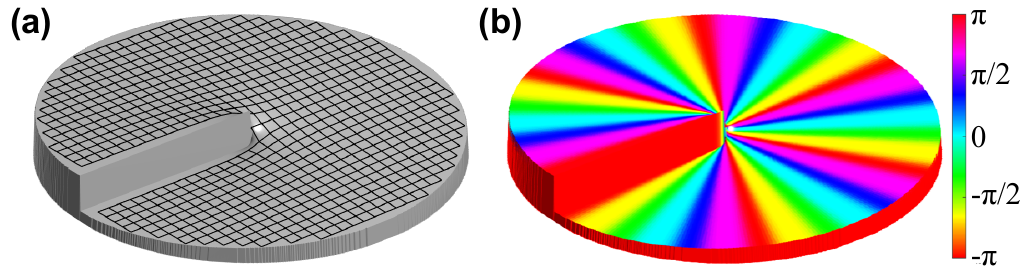}
    \caption{A spiral phase plate. (a)~The surface height, $h(\theta)$. (b)~The corresponding phase shift as a function of position for $l=6$.}
    \label{fig:spp}
\end{figure}

\subsubsection{Single-material spiral phase plate}

Transmissive spiral phase plates, or dielectric slabs whose thickness increases with azimuthal angle, have also been manufactured in several previous studies\cite{Khonina1992,Beijersbergen1994,Oemrawsingh2004,Swartzlander2008,Prasciolu2009,Mari2010,Massari2011,Massari2015}. In this design, the surface height of a transmissive material of refractive index, $n(\lambda)$, satisfies
\begin{equation}
    h(\theta) = \frac{l_0\lambda_0\theta}{2\pi(n(\lambda_0)-1)},
\end{equation}
such that the transmitted beam has the phase shift $l_0 \theta$ at $\lambda_0$ (see Fig.~\ref{fig:spp}). At all other wavelengths, the charge is therefore given by
\begin{equation}
    l(\lambda) = l_0\frac{\lambda_0}{\lambda}\left(\frac{n(\lambda)-1}{n(\lambda_0)-1}\right).
\end{equation}
In the case where the $n(\lambda)$ is approximately constant across the spectral passband, the charge is similar to the spiral phase mirror: $l(\lambda) = l_0 \lambda_0/ \lambda$. 

\subsubsection{Achromatic combinations of spiral phase plates}

We will show in section \ref{sec:achromatic} that the performance of a scalar vortex coronagraph can be improved by optimizing $l(\lambda)$. For instance, using more than one spiral phase plate in series\cite{ChenWang2018} each with different properties provides a means to modify the shape of the $l(\lambda)$ curve. Specifically, for $N$ plates, 
\begin{equation}
    l(\lambda) = \frac{\lambda_0}{\lambda} \sum_j^N l_j \left(\frac{n_j(\lambda)-1}{n_j(\lambda_0)-1}\right),
\end{equation}
where $l_j$ and $n_j(\lambda)$ are respectively the charge and refractive index of the $j$th plate. Here, each plate is individually made of a single material with an arbitrary thickness profile and $l_j$ can take on any positive or negative value. 

Swartzlander (2006)\cite{Swartzlander2006} introduced a spiral phase plate made out of two dielectric materials in contact, where their respective surface profiles are complementary. In that case, the charge relation is given by 
\begin{equation}
    l(\lambda) = l_0\frac{\lambda_0}{\lambda}\left(\frac{n_1(\lambda)-n_2(\lambda)}{n_1(\lambda_0)-n_2(\lambda_0)}\right),
\end{equation}
where $n_1$ and $n_2$ are the refractive indices of each material. The pros and cons of using two-materials in contact versus independent spiral phase plates will be explored in more detail in future work. 

\subsubsection{Pitch multiplicity}

\begin{figure}[t]
    \centering
    \includegraphics[width=\linewidth]{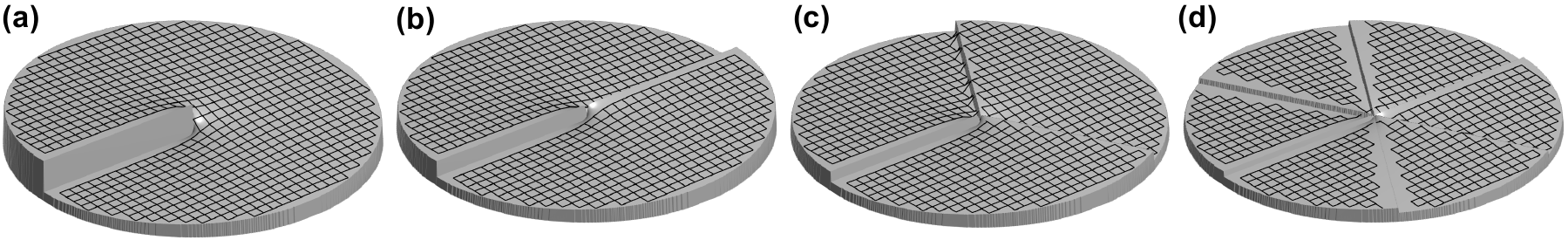}
    \caption{Charge 6 spiral phase plate with pitch multiplicity of (a)~$\mu=1$, (b)~$\mu=2$, (c)~$\mu=3$, and (d)~$\mu=6$. Although the phase shift is theoretically the same in all four cases, the overall thickness is made smaller by a factor of $\mu$.}
    \label{fig:spp_pitch_mult} 
\end{figure}

A spiral phase plate with design charge $l_0$ has an azimuthal phase ramp about the central axis with a phase delay that varies from 0 to 2$\pi l_0$ at $\lambda_0$. However, the ramp can be made up of a combination of ramp sections with discontinuities that are equivalent to an integer number, $\mu$, of $2\pi$~radians. The number of discontinuities in the mask thickness is known as the \textit{pitch multiplicity}\cite{Swartzlander2006}, which can be leveraged to reduce the absolute thickness of a spiral phase plate and the height of the discontinuities (see Fig~\ref{fig:spp_pitch_mult}).

\begin{figure}[t]
    \centering
    \includegraphics[height=0.285\linewidth]{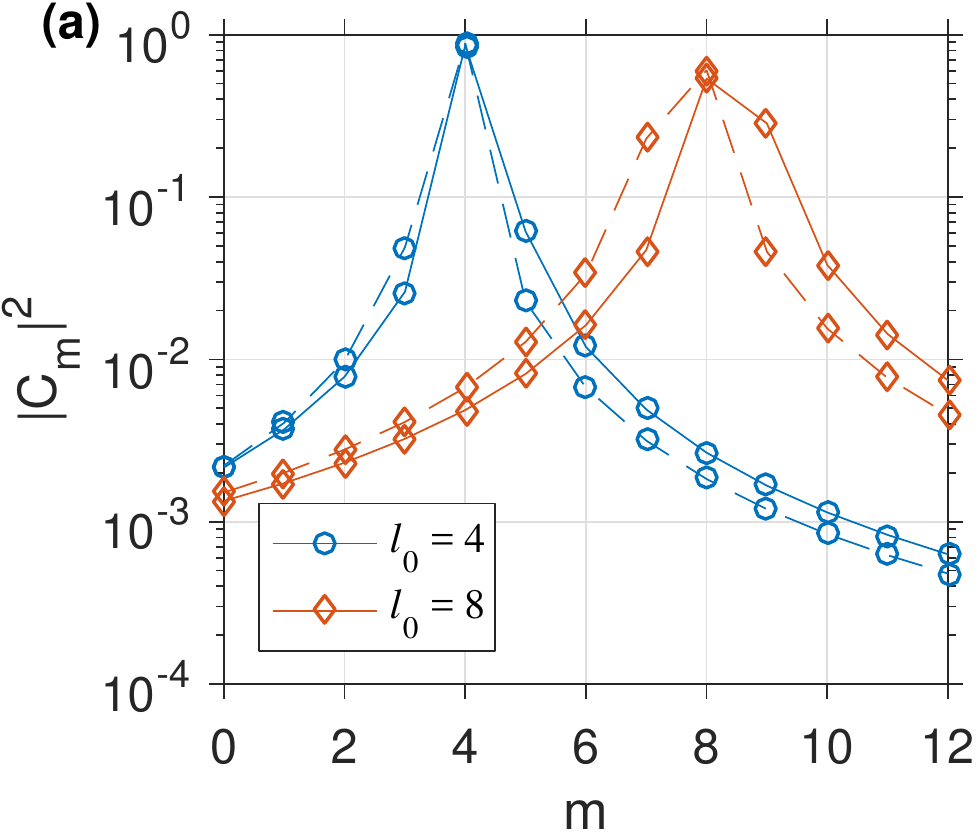}
    \includegraphics[height=0.285\linewidth]{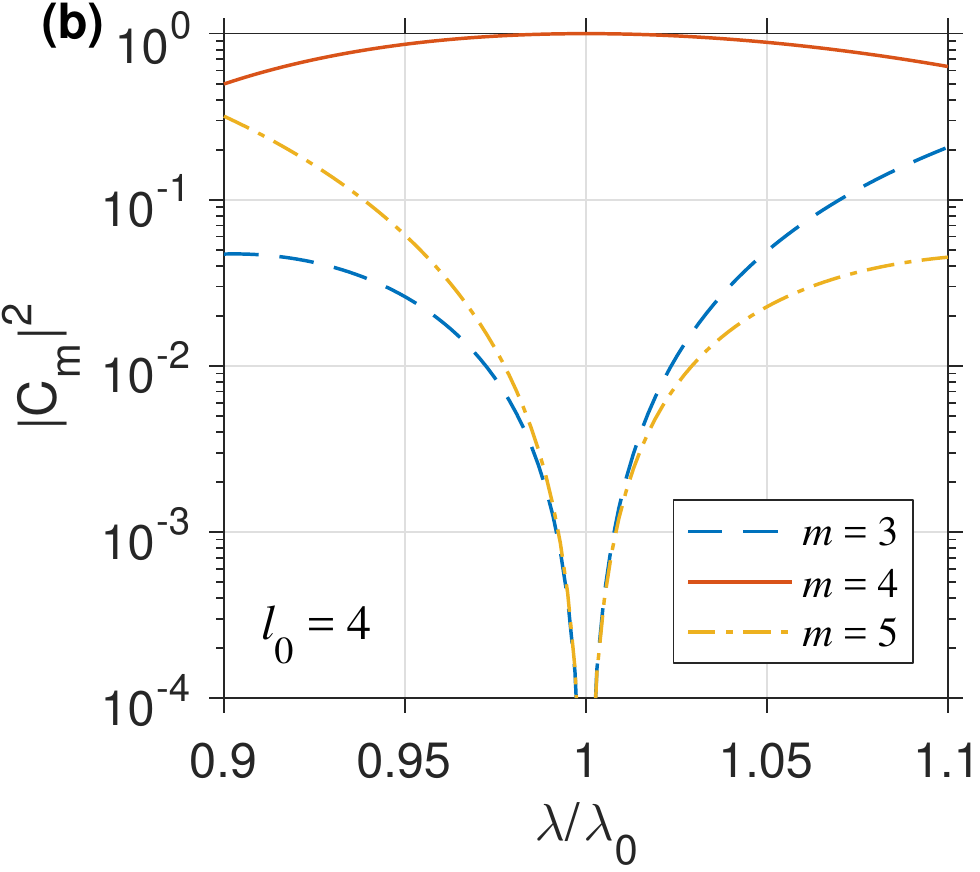}
    \includegraphics[height=0.285\linewidth]{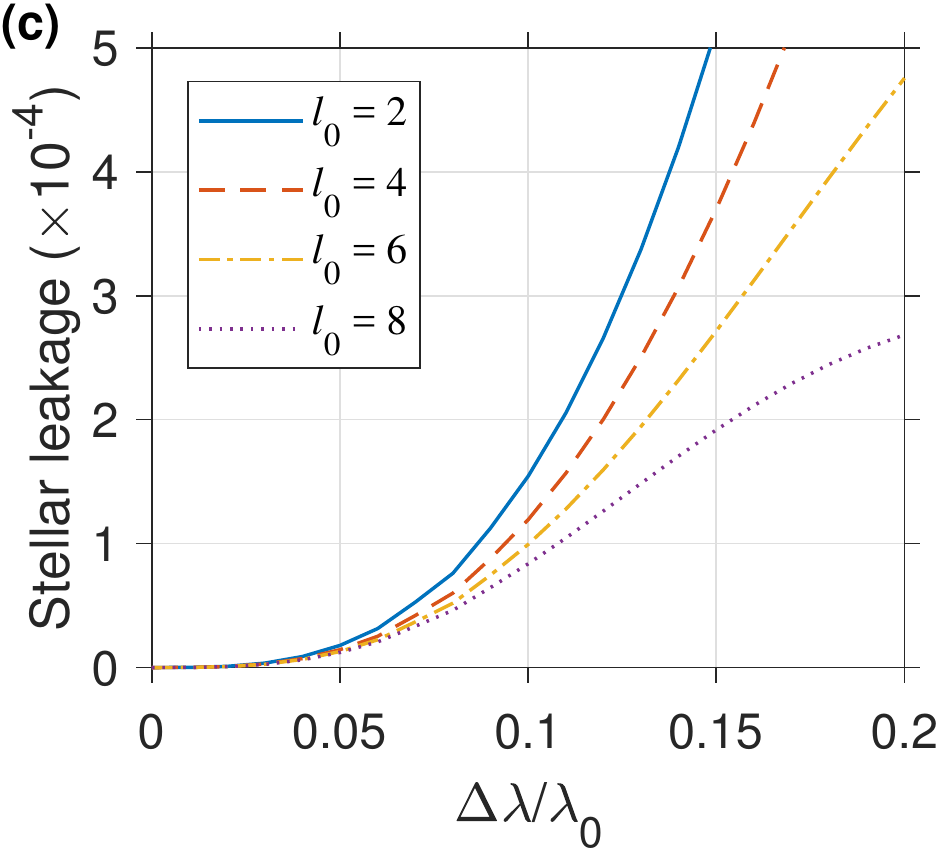}
    \caption{Modal decomposition of a scalar vortex mask with $l(\lambda) = l_0 \lambda_0/ \lambda$. (a)~The vortex spectra for (solid lines)~$\lambda/\lambda_0$=0.95 and (dashed lines)~$\lambda/\lambda_0$=1.05 with design charges 4 and 8. As $\lambda$ deviates from $\lambda_0$, power is transferred to neighboring modes. (b)~The relative power in neighboring modes as a function of wavelength for $l_0$=4. (c)~The fraction of starlight leaked versus the spectral bandwidth for a Lyot stop of radius $b/a=0.95$. }
    \label{fig:cm}
\end{figure}

\subsubsection{The vortex spectrum}

Any optical element that can be fully described by $l(\lambda)$ may also be written as the Fourier series $t(\theta,\lambda)=\sum_m C_{m}(\lambda) e^{im\theta}$, where
\begin{equation}
    C_m(\lambda) = \frac{1}{2\pi}\int_{-\pi}^{\pi}t(\theta,\lambda)e^{-im\theta}d\theta
\end{equation}
is the modal decomposition or so-called vortex spectrum\cite{Swartzlander2005_BBnulling}. Solving for the power in the $m$th azimuthal mode,
\begin{equation}
    |C_m(\lambda)|^2 = \text{sinc}^2\left( l(\lambda)-m \right),
\end{equation}
where we use the normalized sinc function: $\text{sinc}(x) = \sin(\pi x)/(\pi x)$. For instance, assuming we are using a vortex mask with $l(\lambda) = l_0 \lambda_0/ \lambda$, then $|C_m(\lambda)|^2 = \text{sinc}^2\left( l_0 \lambda_0/ \lambda - m \right)$. Figure~\ref{fig:cm}a shows the $|C_m(\lambda)|^2$ distribution as a function of mode for two off-center wavelengths at two example design charges. In general, a scalar vortex optic with design charge $l_0$ puts power into neighboring modes for wavelengths away from $\lambda_0$. Figure~\ref{fig:cm}b shows that the power in the neighboring modes has a strong dependence on the wavelength. 

\subsubsection{Chromatic leakage}

Figure~\ref{fig:cm} illustrates the primary drawback of scalar vortex coronagraphs; i.e. their strong chromatic dependence. As a result of the power entering the neighboring modes when $\lambda$ differs from $\lambda_0$, the central dark region at the second pupil in Fig.~1 begins to fill in with starlight. Since all even and nonzero modes are diffracted completely outside of the pupil, the fraction of starlight that leaks into the dark region as a function of wavelength only depends on $C_0(\lambda)$ and $C_{2k+1}(\lambda)$, where $k$ is an integer. The odd modes are spread both inside and outside of the pupil and each have their own spatial distribution. 
Figure~\ref{fig:cm}c shows the stellar leakage through the Lyot stop over an integrated bandwidth $\Delta\lambda$ about $\lambda_0$. We denote the radii of the dark region at the second pupil and the Lyot stop as $a$ and $b$, respectively. For the calculation in Fig.~\ref{fig:cm}c, $b/a=0.95$. The primary takeaways are that the leakage increases with $(\Delta\lambda/\lambda_0)^2$ and the magnitude of the leakage depends on $l_0$. In later sections, we will show that reducing the size of the Lyot stop, applying wavefront control with deformable mirrors, and using multiple materials for the focal plane mask can improve the starlight rejection of a scalar vortex coronagraph in polychromatic light. 

\subsection{Generalized azimuthal phase functions}

\begin{figure}[t]
    \centering
    \includegraphics[width=\linewidth]{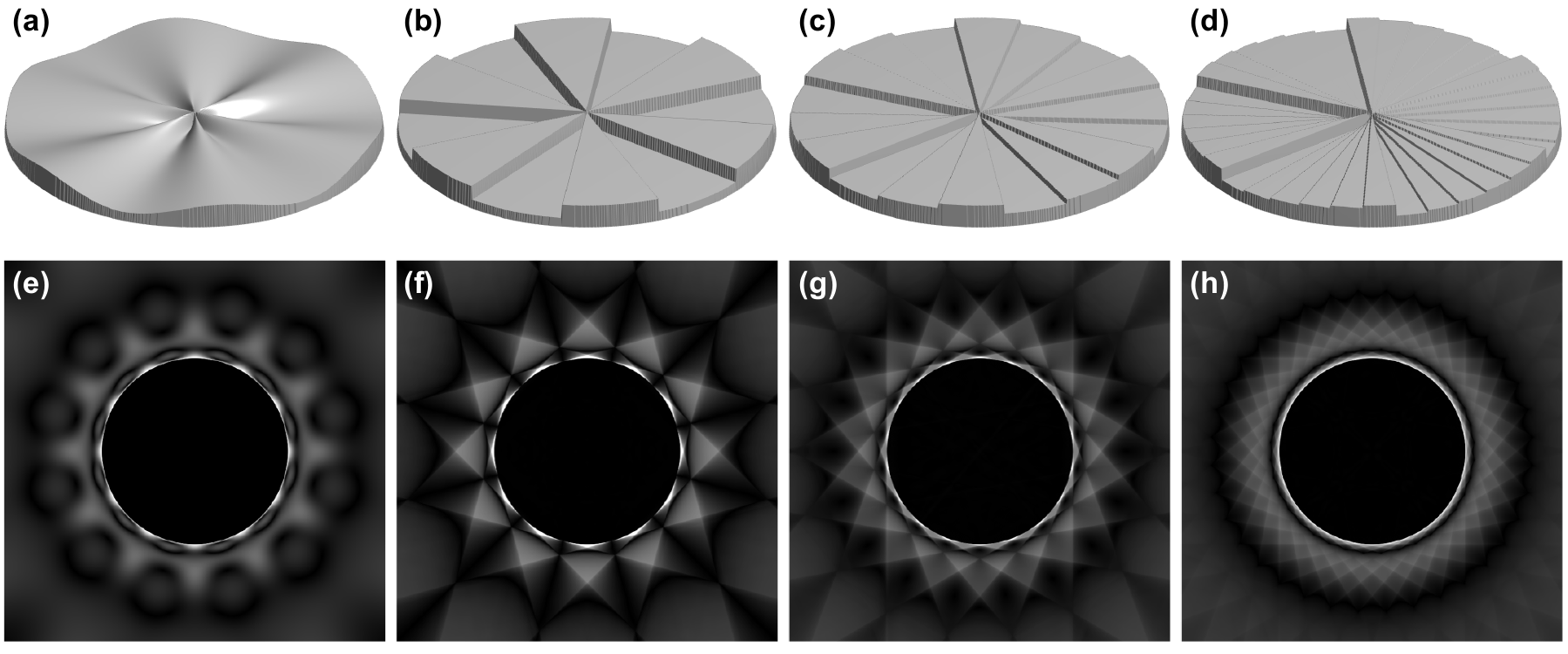}
    \caption{(a)-(d)~Scalar phase masks with similar properties to vortex coronagraphs, including (a)~an azimuthal cosine mask, (b)~a 12-sector mask, (c)~a 3-level staircase mask with $\mu=6$, and (d)~a 6-level staircase mask with $\mu=6$. (e)-(h)~Corresponding field amplitude just before the Lyot stop (``second pupil" in Fig.~\ref{fig:vc}). In theory, all of these mask provide perfect starlight cancellation at $\lambda_0$ and have dominant $m=\pm6$ azimuthal modes. Also see H\'{e}nault (2018)\cite{Henault2018} for an extensive discussion on azimuthal phase masks.}
    \label{fig:azm_phase_masks}
\end{figure}

All of the concepts we introduced for scalar vortex optics generalize to azimuthal phase masks. Ideal starlight cancellation is provided by any mask that is fundamentally a linear combination of vortex modes with even values of $m$ (and $m\ne0$). Figure~\ref{fig:azm_phase_masks} shows examples of sinusoidal, sector, and staircase masks and the corresponding distribution of starlight at the second pupil, where the Lyot stop theoretically blocks all the starlight at $\lambda_0$. The sinusoidal mask has continuously varying thickness, the 12-sector mask has 12 pie-wedge shaped sections that alternate in phase between 0 and $\pi$, and the staircase masks have discretized phase steps as well as pitch multiplicity. 

The performance of the coronagraph, including the planet throughput and low-order aberration sensitivity, is easily predicted from the vortex spectrum. Although the patterns in Fig.~\ref{fig:azm_phase_masks} vary considerably from one another, each have a dominant $m=\pm6$ mode. In fact, the $m=\pm6$ modes are the lowest vortex modes with significant power and therefore the coronagraph is at least as robust to aberrations as a charge 6 vortex coronagraph. This means that they provide first-order rejection of tip/tilt, defocus, astigmatism, coma, and spherical aberrations\cite{Ruane2018_JATIS}. On the other hand, a robust coronagraph tends to have lower throughput at small angular separations and, conversely, a phase mask that has significant power in the low vortex modes (e.g. $m=\pm2$ and  $m=\pm4$) will allow the coronagraph to image closer to the star. 

\begin{figure}[t]
    \centering
    \includegraphics[height=0.33\linewidth]{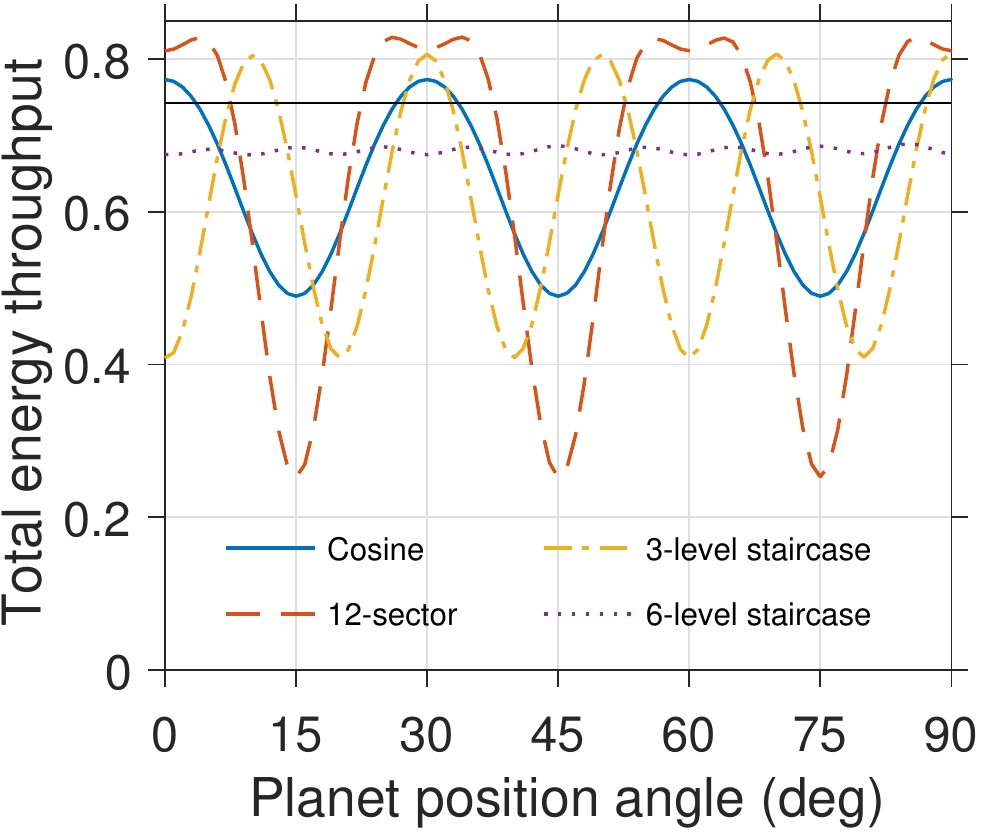}
    \caption{Throughput for a planet at an angular separation of 6~$\lambda/D$ as a function of the planets position angle for the azimuthal phase masks in Fig.~\ref{fig:azm_phase_masks}. The throughput is defined as the fraction of the planet light that passes through a Lyot stop of radius $b/a=0.95$. The solid horizontal black line shows the throughput for a conventional charge 6 vortex coronagraph for comparison.}
    \label{fig:TEthroughput_vs_PA}
\end{figure}

Another important consideration is that steep phase gradients and discontinuities in the azimuthal direction reduce throughput for planets at those position angles. Figure~\ref{fig:TEthroughput_vs_PA} shows the fraction of energy that passes through the Lyot stop ($b/a=0.95$) for a planet 6~$\lambda/D$ from the star as a function of the planets position angle about the star. The 12-sector mask has the largest modulation with throughput ranging from 25\% to 83\% on a period of 30$^\circ$. However, the throughput for the 6-level staircase mask only oscillates between 67\% and 69\% on a 10$^\circ$ period and, therefore, may be more amenable to blindly searching for exoplanets through direct imaging where the planet position angle is unknown. 

The last important feature of the generalized azimuthal phase masks is that some designs may be easier to manufacture than a pure spiral phase pattern. For example, the sector and staircase mask designs can be made with a limited number of etching steps that give a constant depth. To make such masks achromatic, multiple materials can be combined with matching surface patterns, but with their respective depths tuned to reduce leakage into the odd vortex modes across a given passband. 

\section{CORONAGRAPH SYSTEM DESIGN}

In this section, we describe the roles of the other coronagraph components: i.e. the Lyot stop and deformable mirrors. Both provide leverage to reduce chromatic stellar leakage with a scalar vortex coronagraph.

\subsection{The role of the Lyot stop}

\begin{figure}[t]
    \centering
    \includegraphics[height=0.33\linewidth]{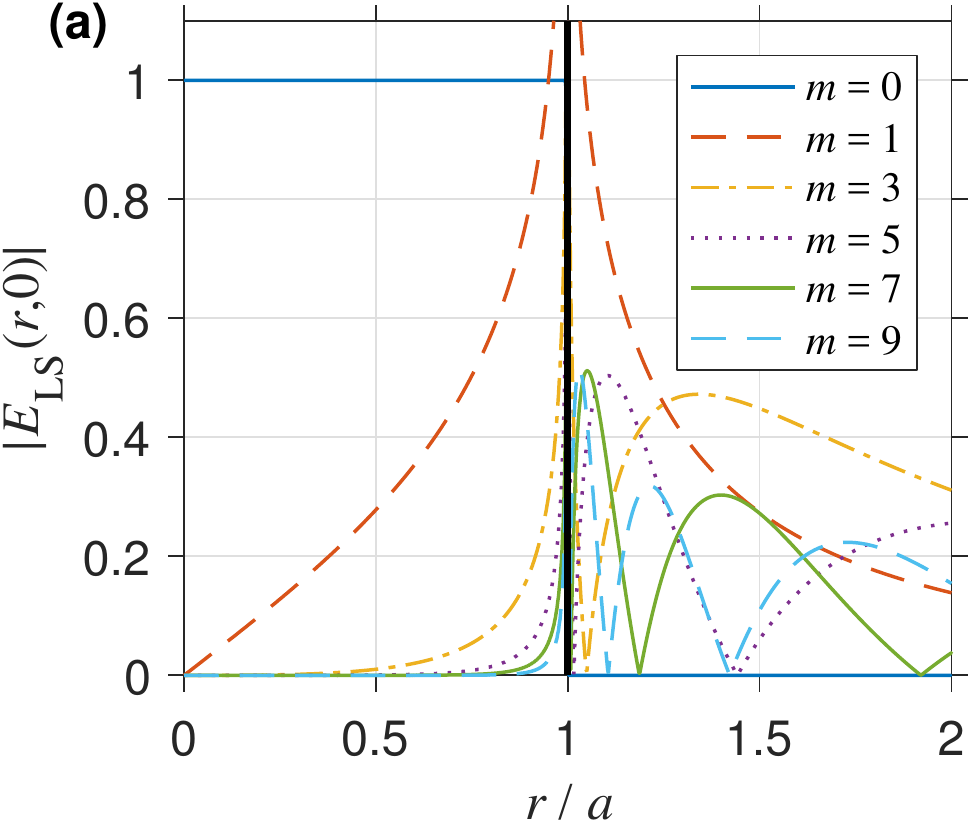}
    \includegraphics[height=0.33\linewidth]{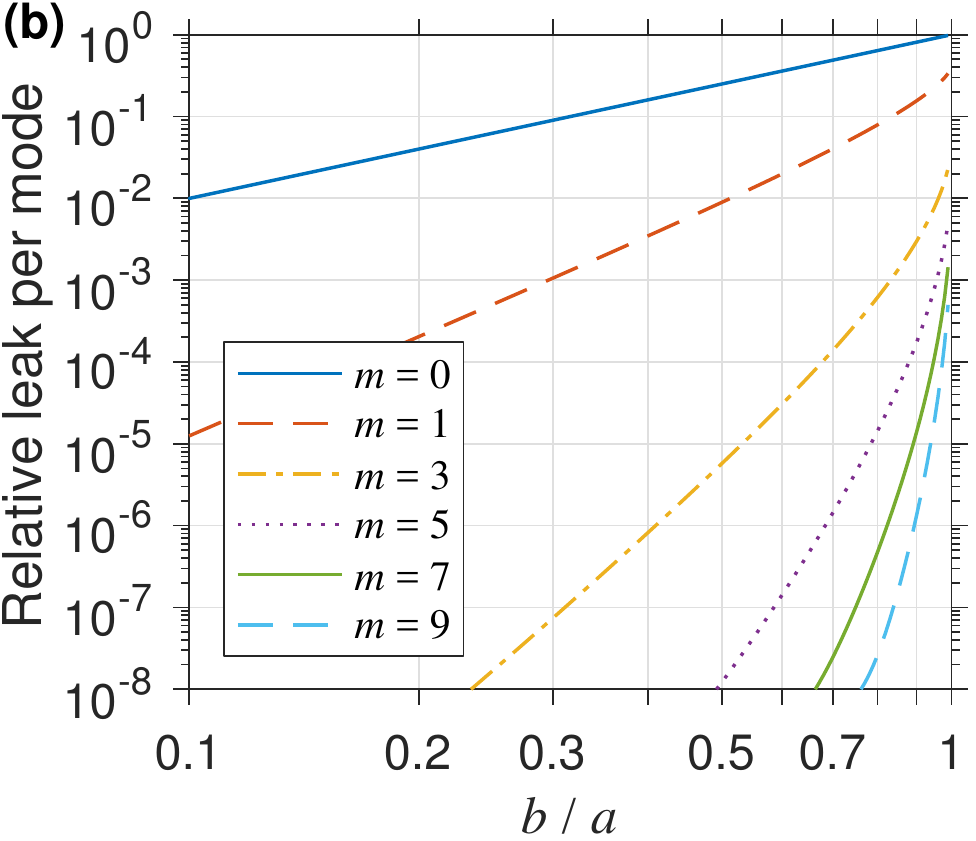}
    \caption{(a)~Field amplitude as a function of radial coordinate in the second pupil (see Fig.~\ref{fig:vc}) for odd vortex modes. The stellar leakage within $r<a$ is more concentrated near the outer edge of the pupil for higher order modes. (b)~Relative stellar leakage per mode when the fractional Lyot stop radius, $b/a$, is reduced. }
    \label{fig:absLS_radpro}
\end{figure}

The radius of the Lyot stop, $b$, is a design parameter that may be used to trade throughput for improved starlight suppression when the vortex mask is imperfect. When $b=a$, the Lyot stop fills the entire dark zone within the second pupil. However, Fig.~\ref{fig:absLS_radpro} shows that reducing the size of the Lyot stop such that $b<a$ blocks a larger fraction of the higher order modes since the leaked light is more concentrated near the outer rim of the dark zone for larger values of $|m|$. Therefore, a coronagraph with higher $l_0$ will have a more substantial improvement in terms of starlight suppression from reducing the Lyot stop radius due to the fact that vortex masks leak most strongly into neighboring modes. For instance, the dominant leakage modes for $l_0=8$ will be $m=7$ and $m=9$; reducing the Lyot stop radius to $b/a=0.8$ respectively suppresses their associated stellar leakage by factors of $<10^{-6}$ and $<10^{-7}$.

\subsection{The role of wavefront control}

Deformable mirrors provide a means to apply chromatic phase shifts in order to apodize the starlight in the final image plane\cite{Trauger2013}. Here, we investigate the expected raw contrast improvement in a scalar vortex coronagraph. We set up a numerical simulation of the coronagraph system with two deformable mirrors, a scalar vortex mask with simple charge dispersion model given by 
\begin{equation}
    l(\lambda) = l_0\frac{\lambda_0}{\lambda},
    \label{eqn:simplemodel}
\end{equation}
and a circular Lyot stop of radius $b$. Using the FALCO software package\cite{Riggs2018,Sidick2018,Coker2018,Ruane2018_FALCO}, we applied Electric Field Conjugation (EFC)\cite{Giveon2007}, to determine the optimal shapes for DM1 and DM2 to minimize the starlight within 2.5-10$\lambda/D$. We varied the charge between $l_0=6$ and $l_0=8$, bandwidth between $\Delta\lambda/\lambda$ of 0.1 and 0.2, and relative Lyot stop diameters $b/a$ of 0.7, 0.8, and 0.9. We also compared solutions for full and half dark holes in the final plane (i.e. 360$^\circ$ and 180$^\circ$ about the star).

\begin{figure}[t]
    \centering
    \includegraphics[width=\linewidth]{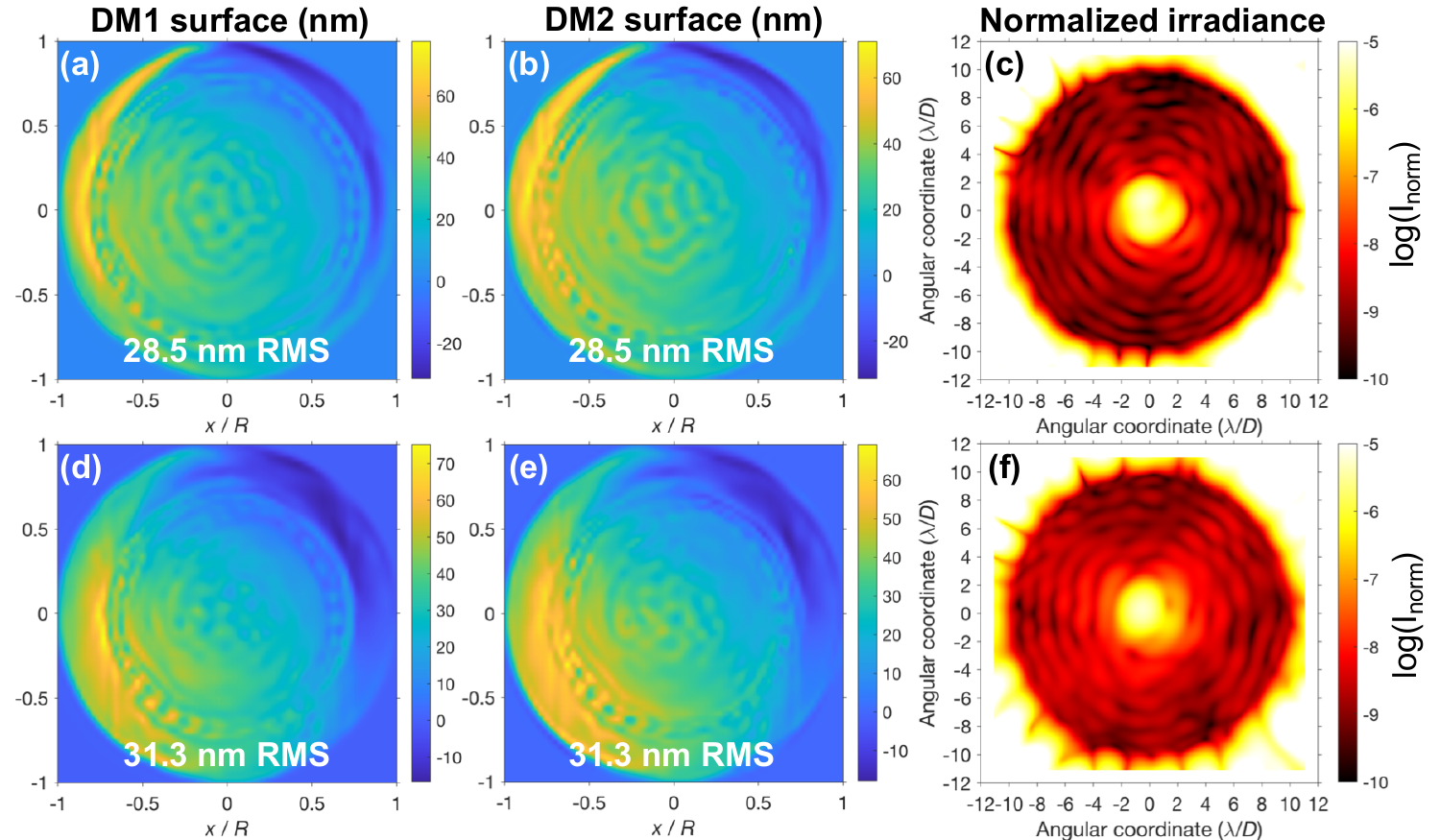}
    \caption{EFC solutions for a scalar vortex coronagraph with $l(\lambda) = l_0\lambda_0/\lambda$. The mirrors shapes for (a)~DM1 and (b)~DM2 give a dark zone within 2.5-10~$\lambda/D$ with normalize irradiance $<10^{-8}$. (a)-(c)~Case 1: solution for $\Delta\lambda/\lambda$=0.1 and $b/a=0.8$. (d)-(f)~Case 2: same as (a)-(c), but with $\Delta\lambda/\lambda$=0.2 and $b/a=0.7$. These examples correspond to the gray rows in Table~\ref{tab:efc_table}.}
    \label{fig:efc_solns}
\end{figure}

\begin{figure}[t]
    \centering
    \includegraphics[width=0.9\linewidth]{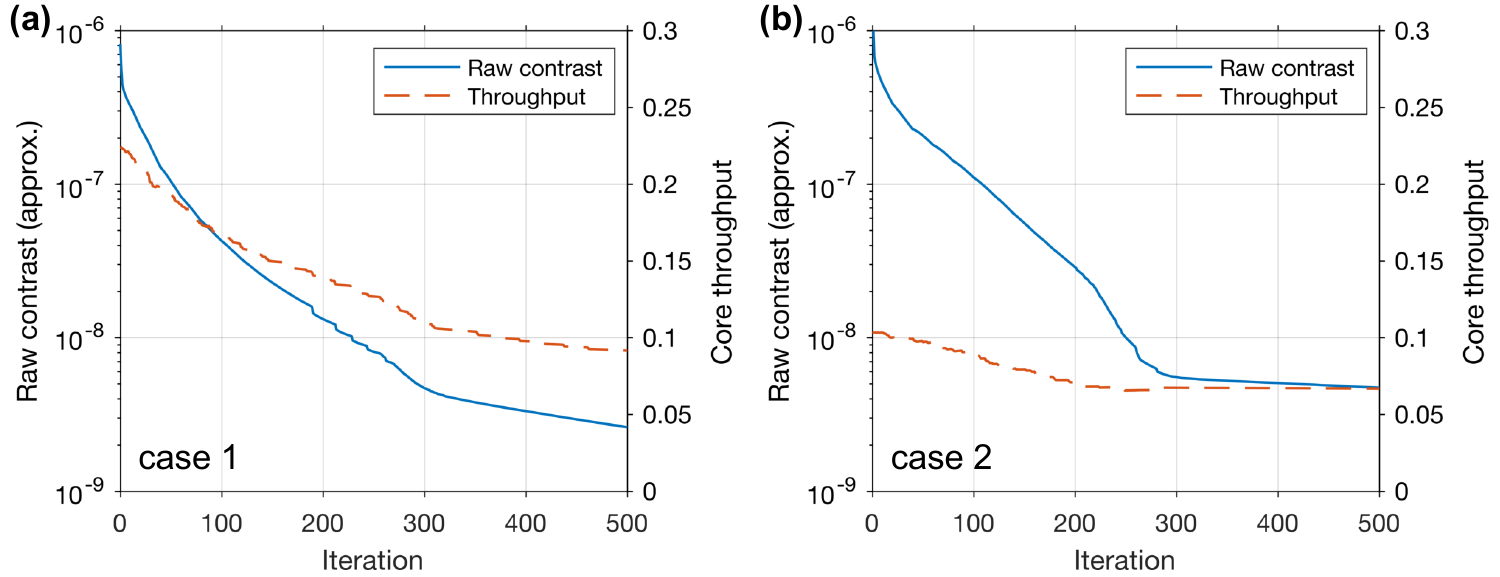}
    \caption{EFC convergence versus iteration for cases (a)~1 and (b)~2. Case 1 has $\Delta\lambda/\lambda$=0.1 and $b/a=0.8$, while case 2 has $\Delta\lambda/\lambda$=0.2 and $b/a=0.7$. Raw contrast is approximated as the irradiance normalized to the peak of the planet PSF at an angular separation of 6~$\lambda/D$. The core throughput is the fraction of planet light within 0.7~$\lambda/D$ of the planet position in the final image. }
    \label{fig:efc_convergence}
\end{figure}

Figure~\ref{fig:efc_solns} shows two example EFC solutions with $l_0=6$ in spectral bandwidths ($\Delta\lambda/\lambda$) of 0.1 and 0.2 (denoted cases 1 and 2, respectively). The surface height on the deformable mirrors are roughly 30~nm rms and the raw contrast is $<10^{-8}$. The EFC convergence plots in Fig.~\ref{fig:efc_convergence} show the tradeoff between raw contrast and the throughput in the planet point spread function (PSF) core defined as within 0.7~$\lambda/D$ of the planet position. The deformable mirror settings that achieve relatively high contrast also distort the off-axis PSF and therefore reduce the encircled energy to 9\% and 7\% in cases 1 and 2, respectively. A full list of solutions is given in Table~\ref{tab:efc_table}. Although we did not find a set of parameters that provides raw contrast of $10^{-9}$ or better for the highly chromatic scalar mask, the simulations show that optimizing the deformable mirrors and Lyot stop can lead to significant performance gains, which should be taken into account when designing a scalar vortex coronagraph. 

\begin{table}[]
    \centering
    \begin{tabular}{|c|c|c|c|c|c|c|c|}
        \hline
        $l_0$ & $\Delta\lambda/\lambda$ & $b/a$ & Dark hole & DM1 surf & DM2 surf & Throughput & Raw contrast \\
        \hline
        6	&	0.1	&	0.9	&	360$^\circ$	&	62.1	&	62.3	&	4.5\%	&	8.3$\times10^{-9}$	\\
        6	&	0.2	&	0.9	&	360$^\circ$	&	71.1	&	71.2	&	3.7\%	&	1.0$\times10^{-8}$	\\ 
        \rowcolor[gray]{.90} 6	&	0.1	&	0.8	&	360$^\circ$	&	28.5	&	28.5	&	9.2\%	&	2.6$\times10^{-9}$	\\ 
        6	&	0.2	&	0.8	&	360$^\circ$	&	33.8	&	34.0	&	7.3\%	&	7.0$\times10^{-9}$	\\
        6	&	0.1	&	0.7	&	360$^\circ$	&	29.8	&	29.8	&	7.7\%	&	3.9$\times10^{-9}$	\\
        \rowcolor[gray]{.90} 6	&	0.2	&	0.7	&	360$^\circ$	&	31.3	&	31.3	&	6.7\%	&	4.7$\times10^{-9}$	\\
        6	&	0.1	&	0.9	&	180$^\circ$	&	28.0	&	28.1	&	20.3\%	&	8.7$\times10^{-9}$	\\
        6	&	0.2	&	0.9	&	180$^\circ$	&	35.0	&	35.3	&	10.9\%	&	1.1$\times10^{-8}$	\\
        6	&	0.1	&	0.8	&	180$^\circ$	&	21.1	&	21.1	&	17.3\%	&	9.6$\times10^{-9}$	\\
        6	&	0.2	&	0.8	&	180$^\circ$	&	26.0	&	26.1	&	9.2\%	&	7.8$\times10^{-9}$	\\
        6	&	0.1	&	0.7	&	180$^\circ$	&	18.1	&	18.1	&	13.0\%	&	1.1$\times10^{-8}$	\\
        6	&	0.2	&	0.7	&	180$^\circ$	&	23.5	&	23.5	&	6.9\%	&	4.5$\times10^{-9}$	\\
        8	&	0.1	&	0.9	&	360$^\circ$	&	26.6	&	27.2	&	10.4\%	&	8.9$\times10^{-9}$	\\
        8	&	0.2	&	0.9	&	360$^\circ$	&	24.1	&	25.2	&	8.0\%	&	1.2$\times10^{-8}$	\\
        8	&	0.1	&	0.8	&	360$^\circ$	&	25.6	&	25.3	&	6.9\%	&	3.3$\times10^{-9}$	\\
        8	&	0.2	&	0.8	&	360$^\circ$	&	20.0	&	20.0	&	7.1\%	&	3.4$\times10^{-9}$	\\
        8	&	0.1	&	0.7	&	360$^\circ$	&	16.6	&	16.6	&	8.0\%	&	3.8$\times10^{-9}$	\\
        8	&	0.2	&	0.7	&	360$^\circ$	&	18.5	&	18.5	&	5.6\%	&	2.4$\times10^{-9}$	\\
        8	&	0.1	&	0.9	&	180$^\circ$	&	19.3	&	19.2	&	12.7\%	&	3.0$\times10^{-9}$	\\
        8	&	0.2	&	0.9	&	180$^\circ$	&	20.8	&	20.9	&	9.3\%	&	4.0$\times10^{-9}$	\\
        8	&	0.1	&	0.8	&	180$^\circ$	&	13.8	&	13.7	&	12.1\%	&	3.0$\times10^{-9}$	\\
        8	&	0.2	&	0.8	&	180$^\circ$	&	16.8	&	16.8	&	8.2\%	&	3.2$\times10^{-9}$	\\
        8	&	0.1	&	0.7	&	180$^\circ$	&	12.3	&	12.2	&	8.8\%	&	3.4$\times10^{-9}$	\\
        8	&	0.2	&	0.7	&	180$^\circ$	&	15.5	&	15.5	&	6.2\%	&	3.6$\times10^{-9}$	\\
        \hline
    \end{tabular}
    \caption{Results of electric field conjugation (EFC) simulations of the simple scalar vortex coronagraph mask model (Eqn.~\ref{eqn:simplemodel}) with the FALCO software package. The relative Lyot stop diameter ($b/a$) is in units of geometric pupil diameter. The dark holes are 3-10~$\lambda/D$ annuli (360$^\circ$) or semi-annuli (180$^\circ$). DM1 and DM2 surfaces are in units of nm RMS. Throughput is defined as the fraction of starlight in the PSF core (area greater than the half maximum). Raw contrast is approximated as the mean irradiance in the dark hole divided by the peak of the off-axis PSF at an angular separation of 6~$\lambda/D$. The gray rows indicate the examples in Figs.~\ref{fig:efc_solns} and \ref{fig:efc_convergence}.}
    \label{tab:efc_table}
\end{table}

\begin{figure}[t]
    \centering
    \includegraphics[width=0.8\linewidth]{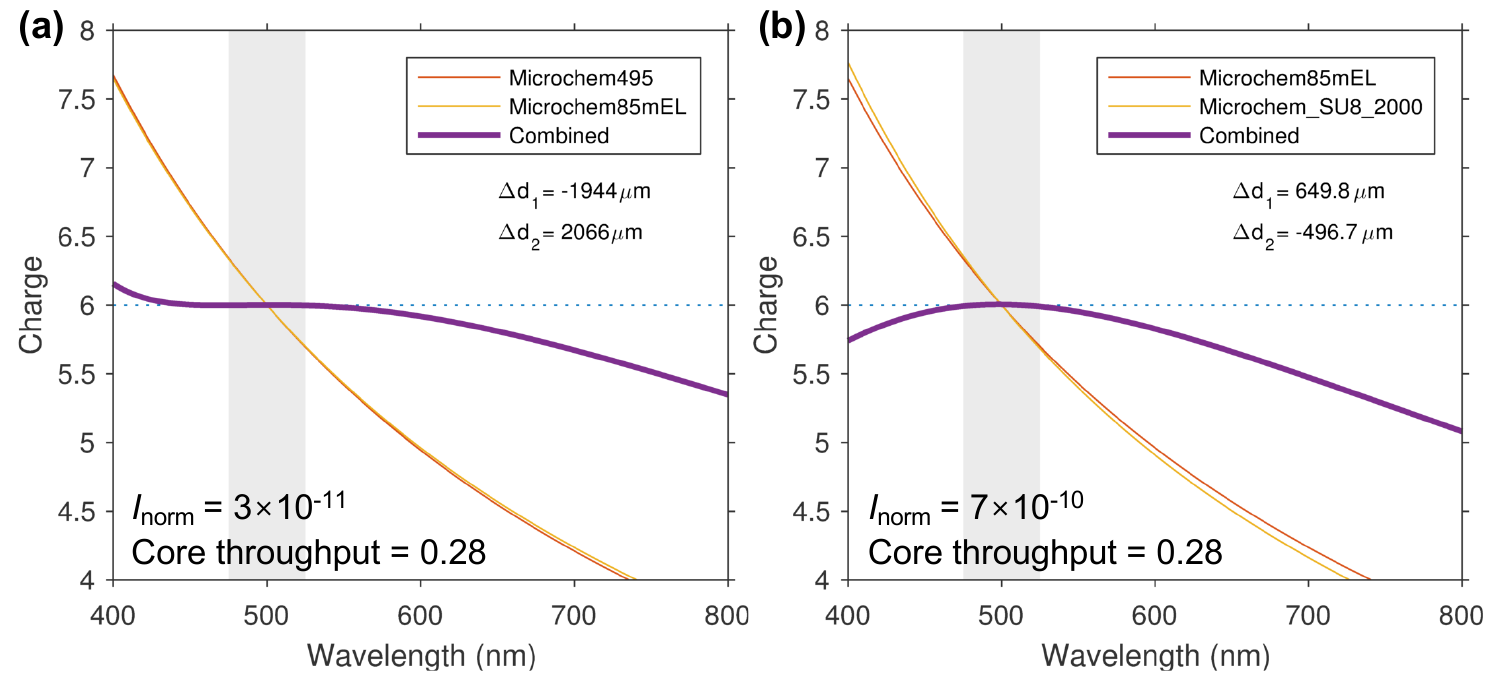}
    \caption{Charge as a function of wavelength, $l(\lambda)$, for two examples of achromatic spiral phase plate combinations with (a)~no thickness constraints and (b)~reduced thicknesses to illustrate the tradeoff between the mask thickness and raw contrast (approximated by the normalized irradiance).}
    \label{fig:achromatic_masks}
\end{figure}

\section{ACHROMATIC SCALAR CORONAGRAPH EXAMPLE}\label{sec:achromatic}

In this section, we show that using two materials to make an achromatic scalar mask can also provide significant gains in coronagraph performance. To demonstrate this, we design two spiral phase plates such that the beam charge is close to 6 over the passband after passing through both of them. Using combinations of photoresists\footnote{\url{http://www.microchem.com/pdf/PMMA_Data_Sheet.pdf}\\ \url{http://www.microchem.com/pdf/SU-8\%203000\%20Data\%20Sheet.pdf}}, we altered the height of each material until the following optimization metric was satisfied: 
\begin{equation}
    \min \int_{\lambda_1}^{\lambda_2} \left| l(\lambda) - l_0 \right|^2 d\lambda,
\end{equation}
where $\lambda_1=\lambda_0-\Delta\lambda/2$ and $\lambda_2=\lambda_0+\Delta\lambda/2$. Figure~\ref{fig:achromatic_masks} shows the resulting $l(\lambda)$ using optimization procedures without and with thickness constraints. We quantify the thicknesses of each plate by $\Delta d_1$ and $\Delta d_2$ where $\Delta d$ represents the step height in a spiral phase plate with $\mu=1$. The sign of $\Delta d$ represents the direction of the phase ramp. 

Compared to using a single material, the advantage of using two materials is that we can achieve better raw contrast and throughput. For instance, in our example in Fig.~\ref{fig:achromatic_masks}a, we used a Lyot stop with $b/a=0.8$ and achieved normalized irradiance of $3\times10^{-11}$ and core throughput of 28\%, which is largely due to the fact that the deformable mirrors are flat. Reducing the allowable $\Delta d$ by approximately 4$\times$ gives normalized irradiance of $7\times10^{-10}$ (see Fig.~\ref{fig:achromatic_masks}b). 

The photoresists used in our example are relatively low index materials which may be driving the design to larger thickness than if we used higher index materials, such as diamond\cite{Karlsson2003}. However, it is also possible to use a constrained optimization to trade the raw contrast for overall thickness. Combining high index materials and using high pitch multiplicity is a promising route to reducing the required thicknesses. We will further explore this and other manufacturing options in future work.  

\section{Conclusions and outlook}

We have provided a theoretical investigation of the use of scalar phase masks in vortex coronagraphs, as an alternative to the more common vector approach. We showed that there are several options for azimuthal phase masks that perform similarly to vortex coronagraphs, which may have some advantages for simplifying manufacturing processes. We identified three ways to improve the bandwidth of scalar vortex coronagraphs: (a)~undersize of the Lyot stop, (b)~use deformable mirrors to reduce chromatic stellar diffraction, and (c)~combine multiple materials to make an achromatic focal plane mask. In future work, we will apply these methods in concert to maximize the coronagraph performance. We will also explore other applications for scalar vortex (or azimuthal phase) masks, such as imaging with ground-based coronagraph instruments\cite{Serabyn2017,Ruane2017,Ruane2019_RDI} and fiber nulling\cite{Ruane2018_VFN, Echeverri2019_VFN, Ruane2019_VFN, Echeverri2019b_VFN} in the near-infrared. 

\appendix

\section{Vector phase masks}\label{sec:appendixVectormasks}

A vector may be represented in terms of Jones matrix $\mathbf{M}$, where 
\begin{equation}
\left[ \begin{matrix}
   U^\prime_x \left(x,y\right) \\
   U^\prime_y \left(x,y\right)  \\
\end{matrix} \right]=\mathbf{M}\left[ \begin{matrix}
   U_x \left(x,y\right)  \\
   U_y \left(x,y\right)  \\
\end{matrix} \right]
\end{equation}
and $U_x \left(x,y\right)$ and $U_y \left(x,y\right)$ are the $x$ and $y$ polarized field components in the $(x,y)$ plane. The polarization transformation provided by an ideal half-wave plate with fast axis orientation angle $\chi$ is
\begin{equation}
\mathbf{M}=\left[ \begin{matrix}
   \cos \chi  & -\sin \chi   \\
   \sin \chi  & \cos \chi   \\
\end{matrix} \right]\left[ \begin{matrix}
   1 & 0  \\
   0 & -1  \\
\end{matrix} \right]\left[ \begin{matrix}
   \cos \chi  & \sin \chi   \\
   -\sin \chi  & \cos \chi   \\
\end{matrix} \right]
=\left[ \begin{matrix}
   \cos 2\chi  & \sin 2\chi   \\
   \sin 2\chi  & -\cos 2\chi   \\
\end{matrix} \right].
\end{equation}
By converting to the circular polarization basis:
\begin{equation}
\mathbf{M}_\circlearrowright=\left[ \begin{matrix}
   1 & i  \\
   1 & -i  \\
\end{matrix} \right] \left[ \begin{matrix}
   \cos 2\chi  & \sin 2\chi   \\
   \sin 2\chi  & -\cos 2\chi   \\
\end{matrix} \right] \left[ \begin{matrix}
   1 & i  \\
   1 & -i  \\
\end{matrix} \right]^{-1}
=\left[ \begin{matrix}
   0 & e^{i2\chi}  \\
   e^{-i2\chi} & 0  \\
\end{matrix} \right].
\end{equation}
A half-waveplate with a spatially-variant fast-axis angle $\chi \left(x,y \right)$ is effectively a phase-only mask: 
\begin{equation}
\left[ \begin{matrix}
   U^\prime_R \left(x,y\right) \\
   U^\prime_L \left(x,y\right)  \\
\end{matrix} \right]=\mathbf{M}_\circlearrowright \left(x,y\right)  \left[ \begin{matrix}
   U_R \left(x,y\right)  \\
   U_L \left(x,y\right)  \\
\end{matrix} \right]
= \left[\begin{matrix}
   0 & {{e}^{i2\chi  \left(x,y\right)}}  \\
   {{e}^{-i2\chi \left(x,y\right) }} & 0  \\
\end{matrix} \right] \left[ \begin{matrix}
   U_R \left(x,y\right)  \\
   U_L \left(x,y\right)  \\
\end{matrix} \right],
\end{equation}
where $U_R$ and $U_L$ are the right- and left-handed circular polarization field components, respectively. Thus, the applied phase function is $\Phi=\pm 2\chi  \left(x,y\right)$, where the sign depends on the handedness of the incident polarization and the phase shift depends only on the local orientation angle of the fast axis $\chi$. A vector vortex mask has $\chi=l\theta/2$, where $l$ is the charge. 

\noindent\textbf{Imperfect retardance in the vector phase mask}

\noindent If the vector phase mask has an imperfect retardance, we can write the Jones matrix as 
\begin{equation}
\mathbf{M}=\left[ \begin{matrix}
   \cos \chi  & -\sin \chi   \\
   \sin \chi  & \cos \chi   \\
\end{matrix} \right]\left[ \begin{matrix}
   1 & 0  \\
   0 & e^{i(\pi+\epsilon_V)}  \\
\end{matrix} \right]\left[ \begin{matrix}
   \cos \chi  & \sin \chi   \\
   -\sin \chi  & \cos \chi   \\
\end{matrix} \right] 
=\left[ \begin{matrix}
   \cos^2\chi-e^{i\epsilon_V}\sin^2\chi  & (1+e^{i\epsilon_V})\cos\chi\sin\chi  \\
   (1+e^{i\epsilon_V})\cos\chi\sin\chi  & \sin^2\chi-e^{i\epsilon_V}\cos^2\chi   \\
\end{matrix} \right],
\end{equation}
where $\epsilon_V$ is the retardance error in the vector phase mask. Converting to the circular polarization basis as above: 
\begin{equation}
\mathbf{M}_\circlearrowright
=\left[ \begin{matrix}
   1-e^{i\epsilon_V} & e^{i2\chi}(1+e^{i\epsilon_V})  \\
   e^{-i2\chi}(1+e^{i\epsilon_V}) & 1-e^{i\epsilon_V}  \\
\end{matrix} \right]
=c_V\left[ \begin{matrix}
   0 & e^{i2\chi}  \\
   e^{-i2\chi} & 0  \\
\end{matrix} \right] + 
c_L\left[ \begin{matrix}
   1 & 0  \\
   0 & 1  \\
\end{matrix} \right],
\end{equation}
where $c_V$ and $c_L$ are constants. The second term results in a stellar leakage whose phase is unchanged by the mask and whose peak is $|c_L|^2$ times fainter than the incident beam, where $|c_L|^2=\sin^2(\epsilon_V/2)$. The fraction of the total power that transfers into optical vortices is $|c_V|^2=\cos^2(\epsilon_V/2)$. For small retardance errors (i.e. $\epsilon_V\ll$1~rad), $|c_L|^2\approx\epsilon_V^2/4$ and $|c_V|^2\approx1-\epsilon_V^2/4$. Generally, $\epsilon_V$ is a function of wavelength and is likely to limit the usable bandwidth of the coronagraph. 

\noindent\textbf{Polarization filtering}

\noindent A method to achieve higher contrast with an imperfect vector phase mask is to block the leakage term by circularly polarizing the beam before it reaches the mask and then using a circular analyzer that only allows the orthogonal polarization to reach the final detector. This will block light the light that does not have the intended phase pattern. To illustrate this, we represent a perfect linear polarizer (LP) and quarter waveplate (QWP) as:
\begin{equation}
\mathbf{J_\text{P}}(\theta)=\left[ \begin{matrix}
   \cos \theta  & -\sin \theta   \\
   \sin \theta  & \cos \theta   \\
\end{matrix} \right]\left[ \begin{matrix}
   1 & 0  \\
   0 & 0  \\
\end{matrix} \right]\left[ \begin{matrix}
   \cos \theta  & \sin \theta   \\
   -\sin \theta  & \cos \theta   \\
\end{matrix} \right]
=\left[ \begin{matrix}
   \cos^2 \theta  & \cos \theta \sin \theta   \\
   \cos \theta \sin \theta  & \sin^2 \theta   \\
\end{matrix} \right],
\end{equation}
\begin{equation}
\mathbf{J_\text{Q}}(\theta)=\left[ \begin{matrix}
   \cos \theta  & -\sin \theta   \\
   \sin \theta  & \cos \theta   \\
\end{matrix} \right]\left[ \begin{matrix}
   1 & 0  \\
   0 & i  \\
\end{matrix} \right]\left[ \begin{matrix}
   \cos \theta  & \sin \theta   \\
   -\sin \theta  & \cos \theta   \\
\end{matrix} \right]
=\left[ \begin{matrix}
   \cos^2 \theta + i \sin^2 \theta  & (1-i)\cos \theta \sin \theta   \\
   (1-i)\cos \theta \sin \theta  & \sin^2 \theta + i \cos^2 \theta   \\
\end{matrix} \right],
\end{equation}
where $\mathbf{J_\text{P}}$ and $\mathbf{J_\text{Q}}$ are the Jones matrices for the LP and QWP, respectively, and $\theta$ is the axis of transmission for the LP and fast axis angle for the QWP. Then, the resulting Jones matrix of the full system in the linear polarization basis is
\begin{equation}
    \mathbf{J_\text{sys}} = \mathbf{J_\text{P}}\left(\frac{\pi}{2}\right) \mathbf{J_\text{Q}}\left(\frac{\pi}{4}\right)
    \left[ \begin{matrix}
    \cos^2\chi-e^{i\epsilon_V}\sin^2\chi  & (1+e^{i\epsilon_V})\cos\chi\sin\chi  \\
    (1+e^{i\epsilon_V})\cos\chi\sin\chi  & \sin^2\chi-e^{i\epsilon_V}\cos^2\chi  \\
    \end{matrix} \right] 
    \mathbf{J_\text{Q}}\left(\frac{-\pi}{4}\right) \mathbf{J_\text{P}}(0) 
    =c_V \left[ \begin{matrix}
    0  & 0   \\
    e^{i2\chi}  & 0   \\
    \end{matrix} \right].
\end{equation}
Thus, the leakage term is blocked while the output beam has the intended phase and is linearly polarized in the $y$-direction.

\noindent\textbf{Polarization filtering with imperfect QWPs}

\noindent In order to determine the effect of a imperfect retardance on the QWP, we introduce both retardance errors $\epsilon_V$ and $\epsilon_Q$ for the vector phase mask and QWPs, respectively. The Jones matrix for the QWP becomes 
\begin{equation}
\begin{split}
\mathbf{\Tilde{J_\text{Q}}}(\theta)&=\left[ \begin{matrix}
   \cos \theta  & -\sin \theta   \\
   \sin \theta  & \cos \theta   \\
\end{matrix} \right]\left[ \begin{matrix}
   1 & 0  \\
   0 & e^{i(\pi/2+\epsilon_Q)}  \\
\end{matrix} \right]\left[ \begin{matrix}
   \cos \theta  & \sin \theta   \\
   -\sin \theta  & \cos \theta   \\
\end{matrix} \right]\\
&=\left[ \begin{matrix}
   \cos^2 \theta + i e^{i\epsilon_Q} \sin^2 \theta  & (1-i e^{i\epsilon_Q})\cos \theta \sin \theta   \\
   (1-i e^{i\epsilon_Q})\cos \theta \sin \theta  & \sin^2 \theta + i e^{i\epsilon_Q} \cos^2 \theta   \\
\end{matrix} \right].
\end{split}
\end{equation}
Repeating the same calculation as above to determine the system Jones matrix gives 
\begin{equation}
    \mathbf{J_\text{sys}} = c_V \left[ \begin{matrix}
    0  & 0   \\
    c_Q e^{i2\chi} + c_Q^\prime e^{-i2\chi}  & 0   \\
    \end{matrix} \right],
\end{equation}
where $c_Q=(e^{i\epsilon_Q}+1)^2$, and $c_Q^\prime=(e^{i\epsilon_Q}-1)^2$. Therefore, even with imperfect retardance in the QWPs, the leakage term is blocked, but the resulting beam is a linear combination of the two conjugate phases (i.e. $\pm2\chi$) and the phase mask does not necessarily impart the intend phase-only pattern. In the case of a vortex coronagraph, there is no additional stellar leakage due to the imperfect QWP retardance since both the positive and negative charges provide starlight cancellation. However, there may be an azimuthally dependent throughput for the planet light. The local transmission of the mask is exactly $T=|c_V|^2\left(3+\cos(2\epsilon_Q)-2 \sin^2(\epsilon_Q)\cos(4\chi)\right)/4$.

\noindent\textbf{Polarization filtering with imperfect LPs}

\noindent For the sake of simplicity, we now assume that the vector phase masks and QWPs are perfect and the LPs have finite extinction ratio, $\gamma^2$. The LP Jones matrix becomes
\begin{equation}
\mathbf{\Tilde{J_\text{P}}}(\theta)=\left[ \begin{matrix}
   \cos \theta  & -\sin \theta   \\
   \sin \theta  & \cos \theta   \\
\end{matrix} \right]\left[ \begin{matrix}
   1 & 0  \\
   0 & 1/\gamma  \\
\end{matrix} \right]\left[ \begin{matrix}
   \cos \theta  & \sin \theta   \\
   -\sin \theta  & \cos \theta   \\
\end{matrix} \right]
=\left[ \begin{matrix}
   \cos^2 \theta + (1/\gamma)\sin^2 \theta   & (1-1/\gamma)\cos \theta \sin \theta   \\
   (1-1/\gamma)\cos \theta \sin \theta  & \sin^2 \theta + (1/\gamma)\cos^2 \theta   \\
\end{matrix} \right]
\end{equation}
and the resulting the system matrix is
\begin{equation}
    \mathbf{J_\text{sys}} = \left[ \begin{matrix}
    0  & -(1/\gamma^2)e^{-i2\chi}  \\
    e^{i2\chi}  & 0   \\
    \end{matrix} \right].
\end{equation}
Next, if we add in a retardance error on the vortex mask, the system matrix becomes
\begin{equation}
    \mathbf{J_\text{sys}} = c_V \left[ \begin{matrix}
    0  & -(1/\gamma^2)e^{-i2\chi}  \\
    e^{i2\chi}  & 0   \\
    \end{matrix} \right] + i \frac{c_L}{\gamma}
    \left[ \begin{matrix}
    1  & 0  \\
    0  & 1  \\
    \end{matrix} \right],
\end{equation}
where the fraction of the beam in the leakage term is $|c_L|^2/\gamma^2=\sin^2(\epsilon_V/2)/\gamma^2\approx\epsilon_V^2/(4\gamma^2)$.

\noindent\textbf{Polarization filtering with imperfect QWPs and LPs}

\noindent Now, we consider imperfections on the QWPs, LPs, and vortex mask. We write the system matrix as
\begin{equation}
    \mathbf{J_\text{sys}} = \left[ \begin{matrix}
    j_{11}  & j_{12}  \\
    j_{21}  & j_{22}   \\
    \end{matrix} \right],
\end{equation}
where
\begin{equation}
    j_{11} = \frac{1}{\gamma}e^{i \epsilon_V/2}e^{i \epsilon_Q}\left( \sin(\epsilon_V/2) - i\cos(\epsilon_V/2) \sin(\epsilon_Q)\cos(2\chi) \right),
\end{equation}
\begin{equation}
    j_{12} = \frac{1}{\gamma^2}e^{i \epsilon_V/2}e^{i \epsilon_Q}\cos(\epsilon_V/2)\left( i\sin(2\chi) - \cos(\epsilon_Q)\cos(2\chi) \right),
\end{equation}
\begin{equation}
    j_{21} = \frac{1}{8}\left(e^{i\epsilon_V}+1 \right)\left( \left(e^{i\epsilon_Q}+1 \right)^2 e^{i2\chi} + \left(e^{i\epsilon_Q}-1 \right)^2 e^{-i2\chi} \right),
\end{equation}
\begin{equation}
    j_{22} = \frac{1}{\gamma}e^{i \epsilon_V/2}e^{i \epsilon_Q}\left( \sin(\epsilon_V/2) + i\cos(\epsilon_V/2) \sin(\epsilon_Q)\cos(2\chi) \right).
\end{equation}
For small retardance errors (i.e. $\epsilon_V\ll$1~rad and $\epsilon_Q\ll$1~rad), 
\begin{equation}
    j_{11} \approx \frac{1}{\gamma}\left(\frac{\epsilon_V}{2} - i\epsilon_Q\cos(2\chi) \right),
\end{equation}
\begin{equation}
    j_{12} \approx -\frac{1}{\gamma^2}e^{-i2\chi}\left(1+i(\epsilon_Q+\epsilon_V/2) \right),
\end{equation}
\begin{equation}
    j_{21} \approx e^{i2\chi}\left(1+i(\epsilon_Q+\epsilon_V/2) \right),
\end{equation}
\begin{equation}
    j_{22} \approx \frac{1}{\gamma}\left(\frac{\epsilon_V}{2} + i\epsilon_Q\cos(2\chi) \right).
\end{equation}
Thus, the leakage term is effectively
\begin{equation}
    |j_{11}|^2=|j_{22}|^2\approx\frac{1}{\gamma^2}\left( \frac{\epsilon_V^2}{4} + \epsilon_Q^2\cos^2(2\chi) \right).
\end{equation}
This implies that having a imperfect retardance on the vector phase mask causes a leakage term that is independent of the fast-axis orientation, $\chi$. However, a retardance error on the QWPs adds a second leakage term with an intensity modulation that depends on $\chi$. The sum of both are reduced by the extinction ratio. For example, if the LPs have an extinction ratio of $\gamma^2=$10$^4$ and the QWPs and vector phase mask have $1^\circ$ retardance error, the leakage is $\sim4\times10^{-8}$ for the worst case $\chi$ orientation.

\acknowledgments  
This work was carried out at the Jet Propulsion Laboratory, California Institute of Technology, under contract with the National Aeronautics and Space Administration (NASA).


\small
\bibliography{Library}   
\bibliographystyle{spiebib}   

\end{document}